\documentclass[a4paper,11pt]{article}
\pdfoutput=1 

\usepackage{jcappub} 

\usepackage[T1]{fontenc} 
\usepackage{hyperref}

\title{\boldmath Interpreting DESI's evidence for evolving dark energy}

 \author{Marina Cortês}
 \author{and Andrew R. Liddle}
 \affiliation{Instituto de Astrof\'{\i}sica e Ci\^{e}ncias do Espa\c{c}o, Universidade de Lisboa,\\ Faculdade de Ci\^{e}ncias, Campo Grande, PT1749-016 Lisboa, Portugal}

\emailAdd{mvcortes@ciencias.ulisboa.pt}
\emailAdd{arliddle@ciencias.ulisboa.pt}

\abstract{The latest results on baryon acoustic oscillations from DESI (Dark Energy Spectroscopic Instrument), when combined with cosmic microwave background and supernova data, show indications of a deviation from a cosmological constant in favour of evolving dark energy. Use of a pivot scale for the equation of state $w$ shows that this evidence is concentrated in the derivative of $w$ rather than its mean offset from $-1$, indicating a new cosmic coincidence where the mean equation of state matches that of the $\Lambda$CDM model precisely in the region probed by the observations. An equivalent way to express this is to say that the dark energy hits the maximum value that it will ever achieve within the observed window. We argue that conclusions on dark energy evolution are strongly driven by the assumed parameter priors and that this coincidence, which we are naming the PhantomX coincidence (where X stands for crossing),  may be a signature of this.}

\begin{document}
\maketitle
\flushbottom

\section{Introduction}
\label{sec:intro}

After over a decade of relentless verification of the standard $\Lambda$CDM model through experiment, cosmology is a field eagerly seeking any signs of deviation from this simple paradigm. Hence the excitement that measurements of baryon acoustic oscillations (BAO) by the Dark Energy Spectroscopic Instrument (DESI) \cite{DESIIII,DESIIV}, when combined with cosmic microwave background (CMB) and supernova observations, show tantalising hints of dark energy evolution \cite{DESIVI}. The deviations within a two-parameter dark energy model are estimated in Ref.~\cite{DESIVI} variously as 2.5 sigma, 3.5 sigma, and 3.9 sigma depending on the supernova dataset included in the compilation (respectively, PantheonPlus \cite{PantheonPlus}, Union3 \cite {Union3}, and DESY5 \cite{DESY5,DESY5data}).

In this article we examine the robustness of this claim, focussing almost entirely on the case of the flat Universe two-parameter equation of state model labelled w0waCDM. 

\section{Dark energy evolution in the w0waCDM model}
\label{sec:w0wa}

\subsection{Models, priors, and pivots}

The w0waCDM model is a phenomenological model of dark energy which features a two-parameter dark energy equation of state \cite{CPL1,CPL2}
\begin{equation}
    w(a) = w_0 + w_a(1-a) \,,
\end{equation}
where $a$ is the scale factor normalized to unity at present, and $w_0$ and $w_a$ are constants. Crucial for our discussion is the choice of prior ranges for these parameters. In Ref.~\cite{DESIVI} they are taken to be uniform in the ranges $[-3,1]$ and $[-3,2]$ respectively, with the additional condition $w_0 + w_a <0$ to allow early matter domination.\footnote{In practice the data go nowhere near this last condition, which would be entirely off the top of the constraint plot shown later.} The cosmological constant, or $\Lambda$CDM, model corresponds to $w_0=-1$ and $w_a = 0$, and a regime in which $w(a)<-1$ is called a `phantom' regime \cite{Caldwell}. 

Theoretical modellers typically find the non-phantom regime much simpler to model than a phantom regime \cite{CST}. For instance, scalar fields with canonical kinetic terms, normally called quintessence models, are necessarily non-phantom, and have already been discussed in the context of the DESI results \cite{newQ1,newQ2} (other physically-motivated scenarios being discussed in Ref.~\cite{Wang}). Ref.~\cite{Wolf} provides a sobering assessment of how useful the $w_0$-$w_a$ parametrization is in developing understanding of the fundamental nature of dark energy. Models that cross between phantom and non-phantom pose additional modelling difficulties due to singularities in the perturbation equations \cite{Vikman,Hu,CaldwellDoran,Nesseris,PFried}, Ref.~\cite{Deffayet} providing a rare example of a model claiming ability to cross this divide without instabilities.

The raw $w_0$--$w_a$ parametrization has the disadvantage that constraints on the two parameters are typically strongly correlated, because there is no useful data constraining the value of $w$ at the present epoch, which is what $w_0$ represents. Fortunately, the model has a reparametrization invariance under a change of pivot scale, i.e.\ the choice of scale factor $a_{\rm p}$ at which to specify $w$; we replace $w_0$ with $w_{\rm p}$ according to
\begin{equation}
w_{\rm p} = w_0 + w_a(1-a_{\rm p}) \,.
\end{equation}
where $a_{\rm p}$ is chosen to decorrelate $w_{\rm p}$ and $w_a$. The pivot scale depends on the data being fit, the main disadvantage of the method being that datasets with different pivots should not be overlaid on the same plot. Marginalized uncertainties on $w_a$ are unchanged by the pivot transformation, but those on $w_{\rm p}$ will be diminished compared to those on $w_0$. The constraint area in the 2D parameter plane is preserved by the transformation, while in a Gaussian approximation the uncertainty on $w_{\rm p}$ will match that of a constant equation of state model ($w_a$ assumed zero).

Pivot scales were originally utilized in CMB studies, capitalising on the freedom to choose a wavenumber $k$ on which to specify the amplitude, slope, etc.\ of perturbations ${\cal P}(k)$. They were first exploited in Ref.~\cite{CGL}, to decorrelate measurements of the spectral index and its running. Later they featured notably in the WMAP Collaboration analyses, which set fixed scales for defining the scalar and tensor amplitudes from its three-year science release onwards \cite{WMAP3}, while optimizing the pivot choice for best power spectrum constraints was demonstrated in Ref.~\cite{CLM}. 

The use of pivots for dark energy, where the pivot is at a time rather than a length, was initiated by Huterer and Turner \cite{HT} who used a pivot within a parametrization of $w$ with redshift. In its modern form using the w0waCDM parametrization, it was popularized by the Dark Energy Task Force Report of Albrecht et al.\ \cite{DETF} who used it in defining a much-used Figure of Merit of observational configurations (see also Refs.~\cite{LinderPivot,MartinPivot}). The pivot scale corresponds to the scale factor at which a given data compilation best constrains the normalization of $w(a)$, which can be considered its average value over the observational window.

\subsection{Evolving dark energy constraints}

We first note that the DESI analysis for models with constant $w$ ($w_a$ assumed zero) finds in all cases results consistent with $\Lambda$CDM \cite{DESIVI}. There is no evidence for a constant offset of $w$ from the $\Lambda$CDM value \cite{DESIVI}.

\begin{figure}[t]
\centering 
\includegraphics[width=.85\textwidth]{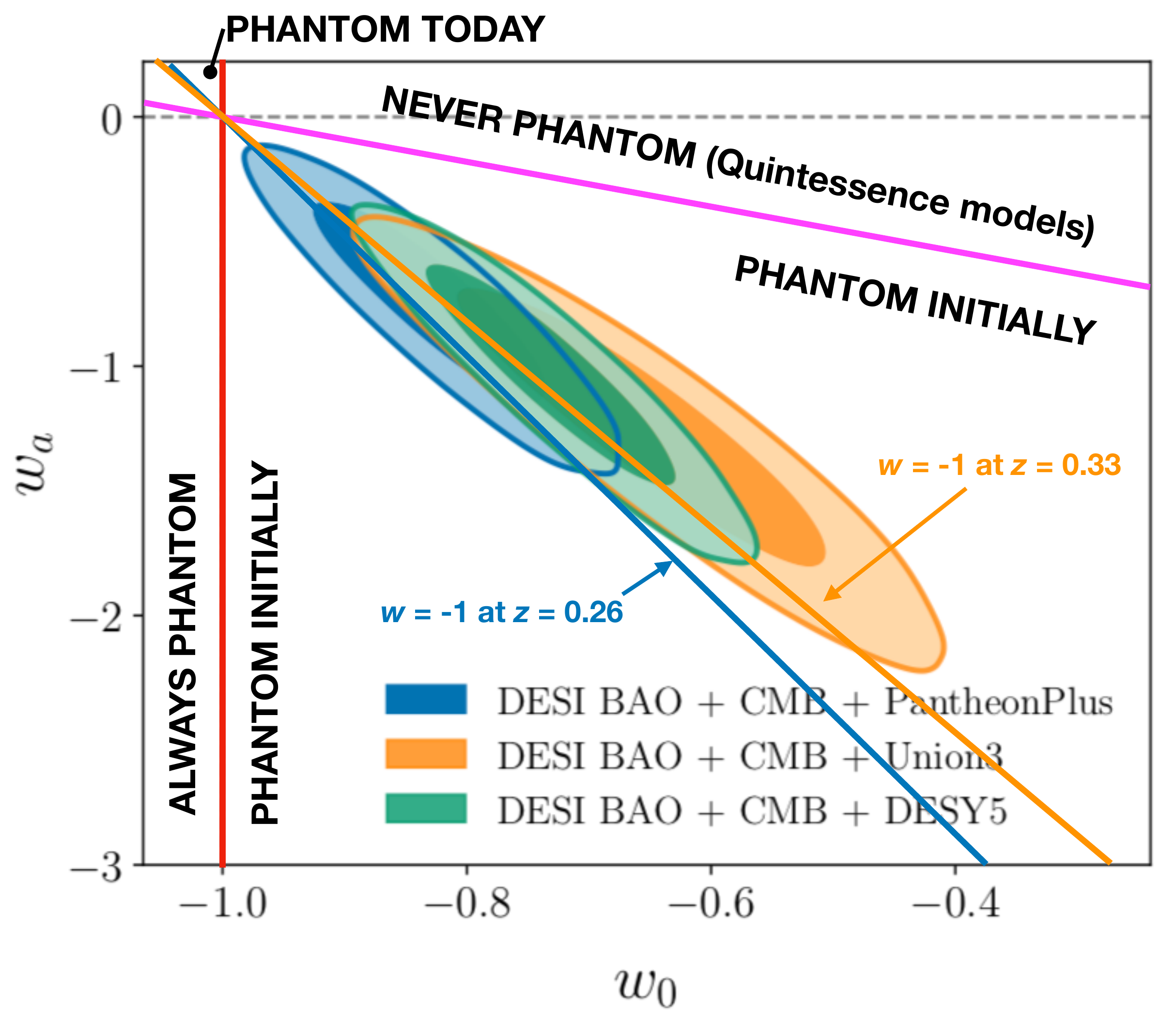}
\caption{\label{fig:DESIpivot} Observational constraints in the $w_0$--$w_a$ plane from Ref.~\cite{DESIVI}, combining DESI BAO and CMB constraints with three different choices of supernova sample. The magenta and red lines partition models into phantom and non-phantom behaviour at early times and today, respectively. In combination they cut the plane into four zones. The blue and orange lines mark parameter values where $w$ crosses $-1$ at redshifts 0.26 and 0.33 respectively. These correspond to the pivot redshifts for the PantheonPlus and DESY5 supernova samples (blue) and Union3 (orange). This shows that all three choices have $w$ close to $-1$ at the pivot scale. [Adapted from Figure 6 of Ref.~\cite{DESIVI}, under Creative Commons BY 4.0 License.]}
\end{figure}

We thus move onto the two-parameter w0waCDM model. In Figure~\ref{fig:DESIpivot} we show the results from the DESI paper (Ref.~\cite{DESIVI}, their Figure 6 right panel), upon which we have superimposed a number of lines to guide our interpretation. The DESI data are BAO measurements from a variety of object types \cite{DESIIII,DESIIV}. The CMB data is mostly from the {\it Planck} Satellite \cite{Planck} with additional information from the Atacama Cosmology Telescope \cite{ACT} in measures of CMB lensing. As described above, three options for catalogues of supernova luminosity--redshift relations are deployed for comparison, PantheonPlus \cite{PantheonPlus}, Union3 \cite{Union3}, and DESY5 \cite{DESY5,DESY5data}.

Before considering the data, note the red and purple lines which divide the parameter plane into four regions. To the lower left are models which are in the phantom regime throughout cosmic history. On the upper right are models that are never phantom, where in particular all quintessence models lie, while the small area in the top left contains models which started non-phantom but transitioned to the phantom regime by the present. All the observational contours lie in the fourth region (the largest region as shown in the figure, though the plot shows only a small part of the entire prior domain); these models transition from an early-time phantom regime to a present-day non-phantom regime.

The most striking feature of the constraints in Figure~\ref{fig:DESIpivot} is that the elongated ellipses point closely in the direction of the $\Lambda$CDM point. This orientation implies that the line defining the pivot scale (the major axis of the ellipses) nearly coincides with the line defining models which  have $w_{\rm p} = -1$. That line is given by $-1-w_0 = w_a(1-a_{\rm p})$ for a chosen $a_{\rm p}$.  The pivot redshifts found in Ref.~\cite{DESIVI} of $z=0.26$ (for PantheonPlus and DESY5) and $z=0.33$ (for Union3) correspond to $a_{\rm p}$ of $0.79$ and $0.75$ respectively, and the lines in model space where $w_p = -1$ at those redshifts are shown by the blue and orange lines. The observational constraints indeed lie on top of those lines and are oriented along them.

In reality the observations are not probing the whole redshift range, but rather a window centred on the pivot scale. That's true even for the CMB probes, which are principally from the dark energy's late-time effect on the angular-diameter distance to last-scattering.  At high redshift the dark energy density is too low to have any observable effect, while at low redshift there is too little volume to take constraining data. 

\begin{figure}[t]
\centering 
\includegraphics[width=.85\textwidth]{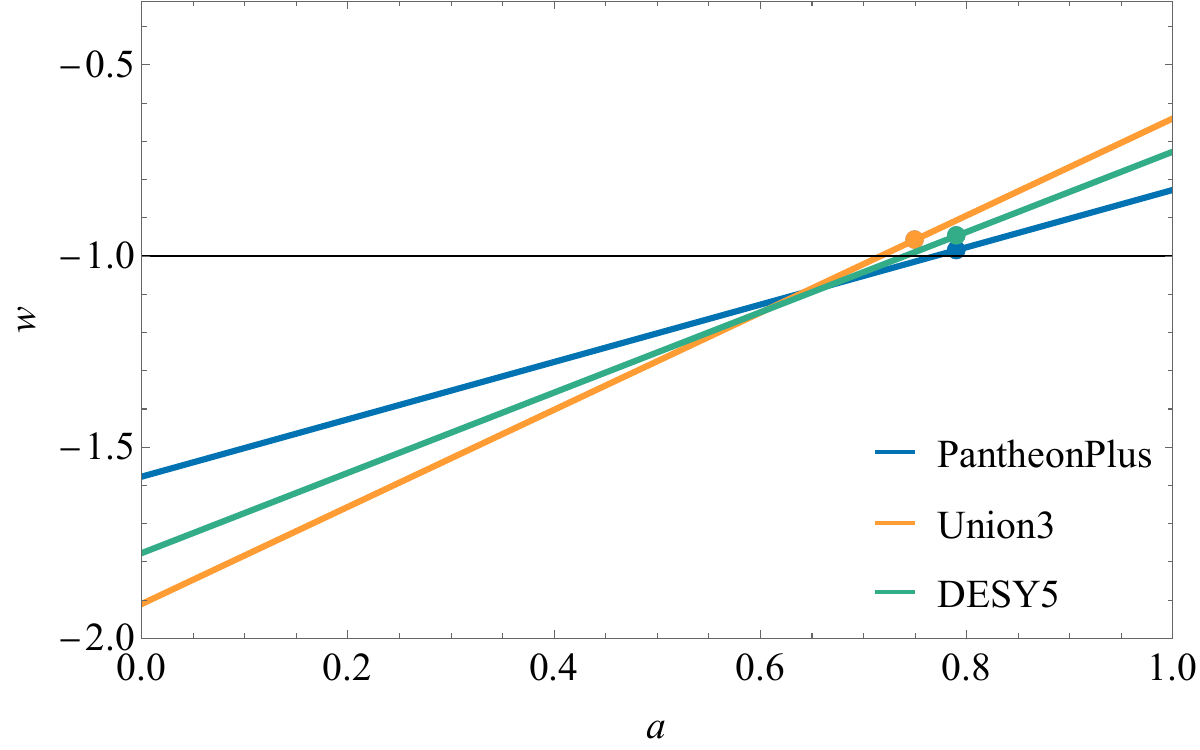}
\caption{\label{fig:wevolution} The best-fit $w(a)$ evolutions for the three choices of supernova dataset, colour coded as in Figure~\ref{fig:DESIpivot}. The pivot scale factors are $a_{\rm p} = 0.79$ for the green and blue lines and $a_{\rm p}=0.75$ for the orange, with the pivot values of $w$ indicated by the blobs. The PhantomX Coincidence is that the blobs are so close to $w=-1$ when most of the evolution is not.
}
\end{figure}

The best-fitting models all have the following characteristic. They start deep in the phantom regime, with $w$ increasing rapidly. Just as we reach the redshifts which are most strongly constrained by the observations, $w$ reaches the $\Lambda$CDM value $w=-1$, to a precision around $\pm 0.02$. At later redshifts, passing outside the observational window, $w$ continues its assumed linear ascent to a value significantly above $-1$. The best-fit $w(a)$ models are shown in Figure~\ref{fig:wevolution}, with the pivot values indicated by blobs.

The outcome is that the preferred regions establish a new cosmic coincidence. Over cosmic history the equation of state of the preferred models exhibits order one variation, but at the epoch of observation it is within the special value of $w=-1$ by a few hundredths. Within the actual range constrained by observations the likely variation is a few tenths in the more extreme models, though there are currently no good estimates of the redshift range that is well constrained by these observations. Gaussian process modelling of $w(a)$ \cite{GPmodels1,GPmodels2} would be one way to achieve this. Since the coincidence is that phantom crossing occurs at or near the centre of the observational window, we call this the PhantomX Coincidence (X stands for crossing). It is separate from the coincidence that dark energy domination ($\Omega_{\rm de} = 0.5$) also takes place during this epoch, the onset of acceleration being somewhat earlier.

Since the phantom transition marks the point at which the dark energy density $\rho_{\rm de}$ stops increasing and starts to decrease, an equivalent statement of the coincidence is that the maximum value that the dark energy density will ever reach happens to lie in the observed window.

Shifting to the pivot scale corresponds to sliding (not rotating) the points on the blue or orange lines of Figure~\ref{fig:DESIpivot} horizontally onto the $w_0 = -1$ axis, dragging the corresponding contours with them. This shifts the contours to have an accurately vertical major axis (corresponding to decorrelation of $w_{\rm p}$ and $w_a$), with all three 95\% contours overlapping the $w_{\rm p} = -1$ axis line.\footnote{The chains used to plot Figure~\ref{fig:DESIpivot} have not yet been made public so we are unable to make the transformation ourselves to replot the figure, though the outcome is easy to estimate by eye.}  This mapping preserves contour areas, i.e.\ the induced prior on $w_{\rm p}$ is still flat.

In a couple of places, Ref.~\cite{DESIVI} refers to a need for extra data to break the degeneracy between $w_0$ and $w_a$. While indeed those parameters are degenerate, this is not really the correct picture. The freedom to make a pivot transformation alone is sufficient to completely remove this degeneracy for any dataset, decorrelating measurements of the amplitude and slope of $w(a)$. What they really mean is that extra data are needed to better constrain $w_a$ in the pivot parametrization.

On a separate point, Figure~1 of Ref.~\cite{DESIVI} shows only a single DESI BAO point, the $z=0.51$ LRG $D_{\rm M}/D_{\rm H}$ ratio of transverse and line-of-sight distances, to be significantly displaced from the best-fit $\Lambda$CDM model. From this one might infer that the significance of the evolution is largely driven by this datapoint. This has seen much discussion since the results were released, especially since the SDSS estimates at similar redshifts \cite{SDSS}, seen in Figure~15 of Ref.~\cite{DESIIII}, do not show this discrepancy from $\Lambda$CDM. However, the analysis in Appendix~A of Ref.~\cite{DESIVI}, which swaps out the low-redshift DESI BAO points for their SDSS equivalents, shows only a modest reduction in the significance of around half a sigma in each of the data combinations. The result is therefore due to a more complex intersection of the different datasets in the full parameter space. Nevertheless, given the significant offset between SDSS and DESI estimates at these redshifts, further data in this regime is crucial.

\subsection{Interpretation}

Let’s look at those initial significances again. Depending on which supernova dataset is combined with DESI BAO and the CMB, there is a claimed discrepancy from $\Lambda$CDM of 2.5 sigma, 3.5 sigma, or 3.9 sigma \cite{DESIVI}, based on a likelihood-ratio comparison. These numbers look so precise, lending credence to the discrepancy. But the interpretation carries a substantial unstated dependence on the choice of prior. 

What motivates that prior? Why is the prior uniform, and with such a wide range? Who thought, before the data came along, that the models with rapidly escalating $w(a)$ at the bottom were as plausible as the quintessence models at the top? That phantom models were {\it a priori} more likely than non-phantom models? Who rates models that have to be patched across the phantom divide to be as reasonable as those which do not? Clearly, the chosen prior is substantially subjective. Certainly it bears little resemblance to the distribution of models considered plausible when the searches for dark energy evolution began in earnest around the time of the Dark Energy Task Force report \cite{DETF}. Our discussion here mirrors our recent discussion of priors in new-physics interpretations of the Hubble tension \cite{CLtension}, where again data is pushing towards models that were considered {\it a priori} unlikely, and in many cases not considered at all, until the data came along.

We could change the prior, for instance by tapering its extension into the deep phantom regime, and in doing so we would lessen this new modest cosmic coincidence that we have described. But if we alter the prior, and downweight  the models that lie furthest from the original theoretical expectations, we will also reduce the discrepancy from $\Lambda$CDM. Since the location that the observations actually constrain lies so close to $w=-1$, the evidence against $\Lambda$CDM would diminish or perhaps disappear entirely. 

While we have concentrated on the flat case, the results allowing varying spatial curvature are extremely similar \cite{DESIVI}. Indeed, curiously the results are all not only consistent with zero curvature, but practically centred upon zero far within the uncertainties. Presuming this to be just another coincidence, our interpretations regarding dark energy apply equal well in the curved case.

 To recap, our principal point is to highlight a so-far-unquantified dependence on model priors. It is clearly discomforting to find the preferred models have a substantially varying $w(a)$ that just happens to be equal (on average, but accurately so) to $-1$ in exactly the range probed by the observations (or, equivalently, that the maximum dark energy density over cosmic history occurred within the observed window). This new cosmic coincidence of phantom crossing in the observed window, which we have called the PhantomX Coincidence, hints that the chosen prior is not a good representation of the likely underlying physics. 

It is not that a particular choice of prior is right or wrong. Bayesian priors are always a matter of judgement, and sufficiently good data will ultimately overturn poor assumptions. Indeed, setting of priors is the one place where the Bayesian method allows the input of personal knowledge and expertise. It's a theorist's chance to show that their reading of Nature is more intuitive than that of their colleagues. But what is imperative is that any conclusions drawn from this kind of analysis must assess the robustness against the large changes that reasonable changes of prior may bring  (see for example Refs.~\cite{JenkinsPeacock,Amendola} for further views on this). In this case, we contend that the robustness has yet to be demonstrated.

\section{Conclusions}

Despite the specific analysis methodology we are highlighting in this manuscript, the first reaction, by any of us in the community, to the release of cosmology results from the first year of DESI’s data, can be nothing other than outstanding and unabashed acclaim. The announcement of this set of articles, from the first ever Stage-IV survey,\footnote{As defined by the Dark Energy Task Force in 2006 \cite{DETF}.} is the result of consistent efforts by a large team of researchers, who have persevered over more than a decade of the survey’s planning, funding acquisition, operational design, and analysis pipeline. 
We also wish to highlight that one of the authors in this work was a member of the DESI survey.\footnote{At the time named BigBOSS.} Specifically we computed and produced the cosmology forecasts obtained through the estimation of both BAO and broadband power spectrum measurements featured in Ref.~\cite{BigBoss}.

We know well that, at the end of the day, every scientist's worst dilemma is the ruthless impartiality with which one must approach scientific procedure. The motivation and revision of one's full set of initial assumptions and convictions is amongst the first priorities in the process of data analysis. It is this scientific rigour that distinguishes our community of enquiry from other disciplines.
Therefore, a more comprehensive understanding of the role of current assumptions in the $w_0$--$w_a$ plane will be vital in fully meeting
the high standards of excellence upheld by the US Department of Energy-led DESI survey collaboration.

~\\
Note added: After our paper was completed, Ref.~\cite{SS} appeared which shows (following Refs.~\cite{LinderPivot,newQ1}) that quintessence models can, despite being everywhere non-phantom, sometimes appear to lie in the phantom region of the $w_0$--$w_a$ plane. They state this is contrary to our current paper in showing that phantom models need to be included. While we agree with this last statement, our paper has never claimed that phantom models should be eliminated entirely, only that they are being given excessive weight in the DESI prior.


\acknowledgments

We thank Cary Sneider for encouragement, Joe Lovano for hospitality during the initial phase of this work, Dragan Huterer and Seshadri Nadathur for clarifications on the DESI analysis, and Luca Amendola and Marco Bruni for comments. This work was supported by the Funda\c{c}\~{a}o para a Ci\^encia e a Tecnologia (FCT) through the research grants UIDB/04434/2020 and UIDP/04434/2020. M.C.\ acknowledges support from the FCT through the Investigador FCT Contract No.\ CEECIND/02581/2018 and POPH/FSE (EC), and A.R.L.\ from the Investigador FCT Contract No.\ CEECIND/02854/2017 and POPH/FSE (EC). M.C.\ and A.R.L.\ are supported by the FCT through the research project PTDC/FIS-AST/0054/2021. This article/publication is based upon work from COST Action CA21136 -- ``Addressing observational tensions in cosmology with systematics and fundamental physics (CosmoVerse)'', supported by COST (European Cooperation in Science and Technology). 



\end{document}